# CCBlock: An Effective Use of Deep Learning for Automatic Diagnosis of COVID-19 Using X-Ray Images

Ali Al-Bawi[1].    Karrar Ali Al-Kaabi[2].    Mohammed Jeryo[3].    Ahmad Al-Fatlawi[4]

**Abstract**

**Propose:** Troubling countries one after another, the COVID-19 pandemic has dramatically affected the health and well-being of the world's population. The disease may continue to persist more extensively due to the increasing number of new cases daily, the rapid spread of the virus, and delay in the PCR analysis results. Therefore, it is necessary to consider developing assistive methods for detecting and diagnosing the COVID-19 to eradicate the spread of the novel coronavirus among people. Based on convolutional neural networks (CNNs), automated detection systems have shown promising results of diagnosing patients with the COVID-19 through radiography; thus, they are introduced as a workable solution to the COVID-19 diagnosis.

**Materials and Methods:** Based on the enhancement of the classical visual geometry group (VGG) network with the convolutional COVID block (*CCBlock*), an efficient screening model was proposed in this study to diagnose and distinguish patients with the COVID-19 from those with pneumonia and the healthy people through radiography. The model testing dataset included 1,828 x-ray images available on public platforms. 310 images were showing confirmed COVID-19 cases, 864 images indicating pneumonia cases, and 654 images showing healthy people.

**Results**: According to the test results, enhancing the classical VGG network with radiography provided the highest diagnosis performance and overall accuracy of 98.52% for two classes as well as accuracy of 95.34% for three classes.

**Conclusions:** According to the results, using the enhanced VGG deep neural network can help radiologists automatically diagnose the COVID-19 through radiography.

1   PhD Student at Ferdowsi University of Mashhad
    ORCID: 0000-0003-0652-7623
    ali.albawi@mail.um.ac.ir

2   Faculty of Veterinary Medicine
    University of Kufa Al-Najaf, Iraq
    PhD Student at Ferdowsi University of Mashhad
    ORCID: 0000-0002-9713-6216
    karrara.hussein@uokufa.edu.iq

3   Faculty of Physical Planning
    University of Kufa Al-Najaf, Iraq
    PhD Student at Ferdowsi University of Mashhad
    ORCID:0000-0003-4695-3491
    mohammedad.hussein@uokufa.edu.iq

4   PhD Student at Ferdowsi University of Mashhad
    ORCID: 0000-0002-0378-8345
    ah.fatlawi@mail.um.ac.ir

# Introduction

In December 2019, the COVID-19 pandemic appeared in Wuhan, China [1-4]. It has adversely been affecting the health and welfare of the world's population and killing many people. It has also impacted the economy of nations where the disease has spread. The novel coronavirus belongs to an outsized group of dangerous viruses [5], which can cause the cold, such as the SARS coronavirus (SARS-CoV). The COVID-19 is also classified as such a diseases. The World Health Organization (WHO) named the infectious disease caused by this type of virus the COVID-19 on Feb 11, 2020 [6]. It has so far been impossible to utterly know this strange virus because its behavior is entirely different. The novel coronavirus is a zoonotic virus because it can be transmitted from animals to humans [7]. This virus is believed to have been passed from bats to humans [8]. The respiratory transmission of the disease among people causes the rapid spread of the pandemic.

The common symptoms of COVID-19 are cough, fever, dyspnea, muscle pain, and fatigue [9]. Causing severe respiratory symptoms, the COVID-19 has increased the intensive care unit (ICU) admission rates. In severer cases, the infection can cause pneumonia, severe acute respiratory syndrome, septic shock, multi-organ failure, and death [7,9]. The primary method of diagnosing the COVID-19 is to conduct a polymerase chain reaction (PCR) procedure [10], which can detect the SARS-CoV-2 RNA from respiratory specimens (collected from the pharyngeal or pharyngeal tracts). In other words, the COVID-19 diagnosis should be confirmed through gene sequencing for respiratory or blood specimens as a critical indicator for the reverse transcription-polymerase chain reaction (RT-PCR) or hospitalization. The PCR test is an essential but susceptible standard; however, it is time-consuming and stressful and can be applied to a limited number of samples. Due to the huge number of infected patients and the abovementioned factors, medical imaging procedures such as the chest X-ray (CXR) and the computerized tomography (CT) scan can play a key role in diagnosing patients with the COVID-19. In fact, radiography examination is fast and easily accessible due to the availability of chest radiology imaging systems in modern healthcare systems.

The main disadvantage of using a CT scan includes the high radiation doses and the costs of scanning. In contrast, conventional radiography or CXR machines are available in hospitals and clinics to produce 2-dimensional (2D) projection images of a patient's thorax. Therefore, it is recommended to use the chest radiography test as a diagnostic method for the COVID-19 [12]. Hence, this study proposes an X-ray imaging technique for potential COVID-19 cases.

The technology of digital image processing has widely been used for medical purposes such as organ segmentation as well as image enhancement and repair to provide the initial support for any subsequent diagnosis [13, 14]. With the rapid development of artificial intelligence (AI), deep learning techniques of automated medical diagnoses have become widely popular with specialists. Deep learning techniques have been used for many medical purposes such as the breast cancer diagnosis [15], classification of brain diseases [16], and diagnosis of pneumonia [17]. With the outbreak of the COVID-19 pandemic and the disproportionate number of patients to the preparation of diagnostic medical staff, artificial intelligence researchers must employ their competencies to detect this disease and mitigate its spread.

Recently, numerous studies have suggested the automated diagnosis of COVID-19. This section reviews some of the related studies, the results of which are discussed later. Ioannis *et al.* [18] proposed the use of transfer learning with a deep model to diagnose the COVID-19. Their model showed good accuracy in the categorization of two classes and three classes. Tulin *et al.* [19] introduced a DarkNet model as a classifier. Their proposed model consisted of 17 convolutional layers with different filters. Linda Wang *et al.* [20] proposed a deep convolutional neural network, named the COVID-Net, by adopting a human-machine collaborative design strategy. They also collected a dataset of 13800 chest x-ray images called the COVIDx. Hamdan *et al.* [21] presented the COVIDX-NET framework based on seven deep neural networks such as VGG-19, DenseNet121, and ResNetV2 to train the dataset and diagnose the COVID-19. Ali Narin *et al.* [22] employed the pre-trained convolutional neural network-based models (*e.g.* ResNet50, InceptionV3, and Inception ResNetV2) to predict a small dataset. The study by Prabira Kumar Sethy *et al.* [23] differs from the abovementioned studies in terms of the research strategy because they first extracted the deep features through a deep convolutional network (ResNet50) and then classified the COVID-19 cases based on the remaining chest x-ray images by using a support vector machine (SVM). Due to the data insufficiency, Pedro Bassi *et al.* [24] adopted the transfer learning strategy and developed a deep neural network (CheXNet) which was pre-trained with images of 14 chest diseases. CheXNet is an extension of DenseNet121 trained on ImageNet and retrained in 14 classes of chest x-rays.

There are also many other studies of the COVID-19 diagnosis based on deep neural networks with CT scan images [25-29].

However, deep learning techniques are still used rarely to diagnose the COVID-19 in X-rays, although they produce reasonable accuracy. Due to the need for the quicker interpretation of radiography images, this study proposes an automated technique for distinguishing the COVID-19 cases from patients with pneumonia and the healthy people by enhancing the classical VGG network with radiography.

The next section discusses the methodology for developing the proposed network through transfer learning and the general clarification of deep convolutional neural networks used in the VGG network enhancement architecture. *Experiments and Results* give a general review of the dataset used in experiments and present the results of experiments conducted to evaluate the efficacy of the proposed VGG network in comparison with the previous related works. The *discussion* addresses the experiment results and research limitations. Finally, the *conclusion* draws the main conclusions and states the research rationale.

## Materials and Methods

### Dataset

Sufficient data must be available to develop and improve a diagnostic tool. To overcome the paucity of X-rays related to the COVID-19 cases, three different open sources were employed to collect a sufficient number of X-rays to train and test the proposed network. The research dataset includes the human chest X-rays taken by a widely available radiography machine. A challenge to network training is the imbalance of data; therefore, the dataset preparation was balanced. For this purpose, 1828 chest X-rays were selected. The first of the three sources used in the study came from Dr. Cohen, who collected data from public sources that did not violate patient privacy.

Then 241 chest X-rays of the COVID-19 patients were extracted from Dr. Cohen's dataset [30]. The images showed different angles including the posteroanterior (PA), anteroposterior (AP), laying down (AP supine), and lateral (L) views. Moreover, some images showed the same patient from different angles but at different offset days of the disease, whereas other images indicated different patients. The benefits and impacts of this method will be discussed in the *discussion*.

The second source used in this study came from the Kaggle platform. It consisted of 79 chest X-rays of COVID-19 patients [31]. There are ten similar images between the two sources; thus, they were deleted to avoid duplication. As a result, there were 310 chest X-rays of the COVID-19 patients in total. The third source also came from the Kaggle platform. It contained a broad set of chest X-rays for patients with pneumonia and healthy people [32]. There were 864 chest X-rays of the patients with pneumonia: 467 images of bacterial pneumonia and 397 images of viral pneumonia. There were also 654 chest X-rays of the healthy people in the same dataset, which was divided into two sections, *i.e.* the training set (27%) and the testing set (73%) as shown in Fig.1 and Table 1.

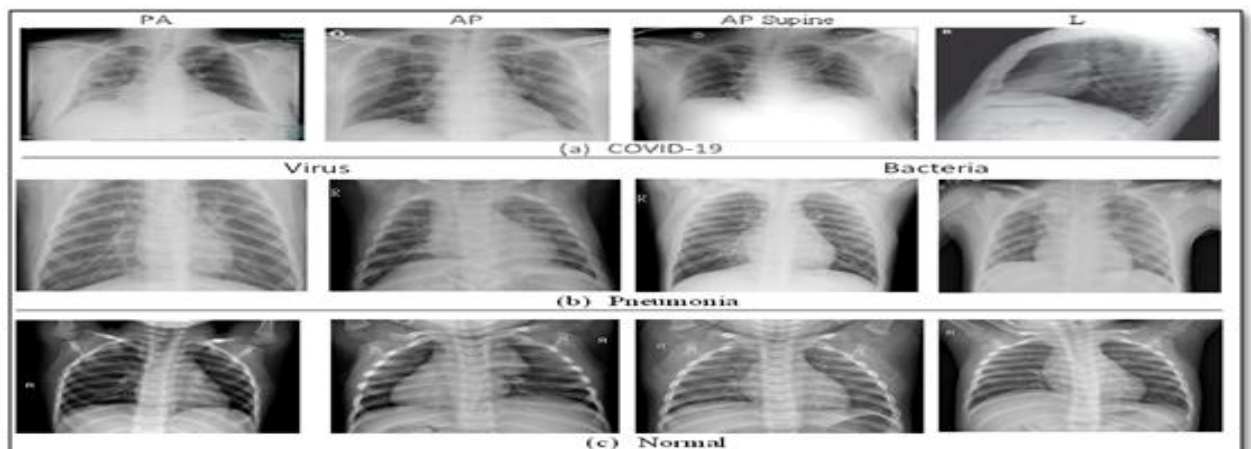

**Fig. 1**: Samples of chest X-rays; (a) the COVID-19 cases from four different angles; (b) viral and bacterial pneumonia cases; and (c) healthy people

Table 1: The dataset details

|              | COVID-19 | Pneumonia (Virus Bacterial) | Normal |
|--------------|----------|------------------------------|--------|
| **Train**    | 84       | 233                          | 176    |
| **Test**     | 226      | 631                          | 478    |
| **Train + Test** | 310  | 864                          | 654    |

**Transfer Learning**

Transfer learning is a strategy for transferring the knowledge extracted by a neural network from specific data to solve a problem. However, it is applied to a new task including new and usually insufficient data to train neural networks from the beginning [33].

In deep learning, it is necessary to access large and sufficient amounts of data for the proper training of neural networks, as data availability for initial training is an essential factor for the successful implementation of the CNN training process in extracting the distinct features images. Regarding the insufficient amount of training data as in medical images, there is no choice but to resort to the ability of neural networks trained in a sufficient dataset to extract the essential features of images. This process is called the transfer of learning. There are two strategies for transferring learning.

The first strategy is to use neural networks to extract important features from data while retaining the trained network architecture where the trained network outputs are the data features given to the classifier network [34]. In the second strategy, the network architecture is adjusted to use its pre-trained weights attached to a parallel architecture containing untrained weights that were trained through the available data used in this study. The most popular neural networks that employ transfer learning for medical purposes are ImageNet-trained networks used in the ImageNet Large-Scale Visual Recognition Challenge (ILSVRC) [35]. The networks trained in this dataset for medical purposes include VGG-16, VGG-19, and Res-Net.

**Deep Learning Classifiers**

This section describes deep classification networks used in experiments.
VGG-Net: This network was developed by K. Simonyan and A. Zisserman in 2014 [36] for ILSVRC-2014. It performed well in the ImageNet data classification. There are two versions of this network architecture differing in depth. The first is called VGG-16, which contains 13 convolution layers, three fully connected layers, and five pooling layers. The second is called VGG-19, which contains 16 convolution layers, three fully connected layers, and five pooling layers.

**The Proposed Model**

Deep learning has brought about a breakthrough in different areas of artificial intelligence such as detection and identification of images, people, and sounds. The word "deep" indicates an increase in the number of layers. A model that uses one or more hidden layers is called a deep model. These are called CNN models, *i.e.* a convolutional neural network, in which the word "convolutional" denotes the presence of convolutional layers that contain a set of weighted filters trained through learning data. An advantage of CNN is to extract the input features. Another important layer is the fully connected layer located at the end of the network. This layer contains a set of weights trained through training data in the training phase. There are also other layers called activity layers that include the non-linear activity layer (ReLu) aiming to delete negative values.

Neural networks are trained by using a set of enhancers, the most important of which is the stochastic gradient descent (SGD), which operates at the expense of the error derivative, for use in the process of updating the weights in the layers of a deep model as in the convolutional layers and fully connected layers. The training process includes updating weights with a dependent error derivative and a small learning rate. Training a convolutional neural network requires large amounts of data, which are not usually available in medical diagnostic tasks. Despite the rapid and widespread prevalence and a large number of patients with the COVID-19 that, it is impossible to collect enough data to train neural networks from the beginning due to the harsh conditions of the pandemic. Therefore, there is no choice but to resort to the scant data existing on different platforms.

It is unnecessary to build a deep model from the beginning because pre-built models (VGG-16, VGG-19, and ResNet) can be modified and used. VGG-16 is a network with low computational complexity due to the small dimensions of its filters. This network contains 16 learnable layers that have 9-pixel filters; thus, it runs fast due to the smallness of its filters. A transfer learning method is to add learnable layers at the end of the network to train it in the current data. This will improve network learning in terms of accurate classification. Therefore, three learnable convolutional layers were added and trained by using the previously introduced data to improve the network performance and obtain the expected results.

In this study, VGG-16 was selected and modified by adding three convolutional layers, each of which was followed by the ReLu-Batch normalization layer. This modification will improve the classification process due to the presence of several untrained filters that will be trained by using the training data.

The added convolutional layers have different numbers of filters (512,256,128). This model resembles VGG-19 in the number of learnable layers. In fact, VGG-19 has 19 learnable layers containing the weights that are trained by using training data. The original model (VGG-16) contains 16 learnable layers. In this study, one of the fully connected layers was deleted from the basic model; therefore, 15 layers remained. The proposed model is similar to VGG-19 in terms of the number of convolutional layers. These three added layers were named the *Convolutional COVID Block (CCBlock)*.

As discussed earlier, it is possible to use pre-trained networks in different strategies. In this study, the second proposed strategy allowed the use of pre-trained networks as a part of the deep model in which the part added to the model was trained through the training data. Table 2 shows the layers used for the proposed model with all relevant features, whereas Fig. 2 shows the procedure for the proposed model CCBlock (training and testing phases)

**Table 2:** The layers and their features in the proposed model (for two or three categories)

|       | Layer         | Feature map | Size        | Trainable | Pre-Trained |
|-------|---------------|-------------|-------------|-----------|-------------|
| input | Image         | 1           | 224x224x3   | False     | False       |
| 1     | 2xConvolution | 64          | 224x224x64  | True      | True        |
| 2     | Maxpooling    | 64          | 112x112x64  | False     | False       |
| 3     | 2xConvolution | 128         | 112x112x128 | True      | True        |
| 4     | Maxpooling    | 128         | 56x56x128   | False     | False       |
| 5     | 2xConvolution | 256         | 56x56x256   | True      | True        |
| 6     | Maxpooling    | 256         | 28x28x256   | False     | False       |
| 7     | 3xConvolution | 512         | 28x28x512   | True      | True        |
| 8     | Maxpooling    | 512         | 14x14x512   | False     | False       |
| 9     | 3xConvolution | 512         | 14x14x512   | True      | True        |
| 10    | Maxpooling    | 512         | 7x7x512     | False     | False       |
| 11    | **1xConvolution** | 512     | 5x5x512     | True      | False       |
| 12 | **BatchNorm** | 512 | 5x5x512 | True | False |
| 13 | **1xConvolution** | 256 | 3x3x256 | True | False |
| 14 | **BatchNorm** | 256 | 3x3x256 | True | False |
| 15 | **1xConvolution** | 128 | 1x1x128 | True | False |
| 16 | **BatchNorm** | 128 | 1x1x128 | True | False |
| 17 | Flatten | 128 | 1x128 | False | False |
| 18 | FC | - | 1x256 | True | False |
| 19 | FC+Softmax | - | 1x3 or 1x2 | True | False |

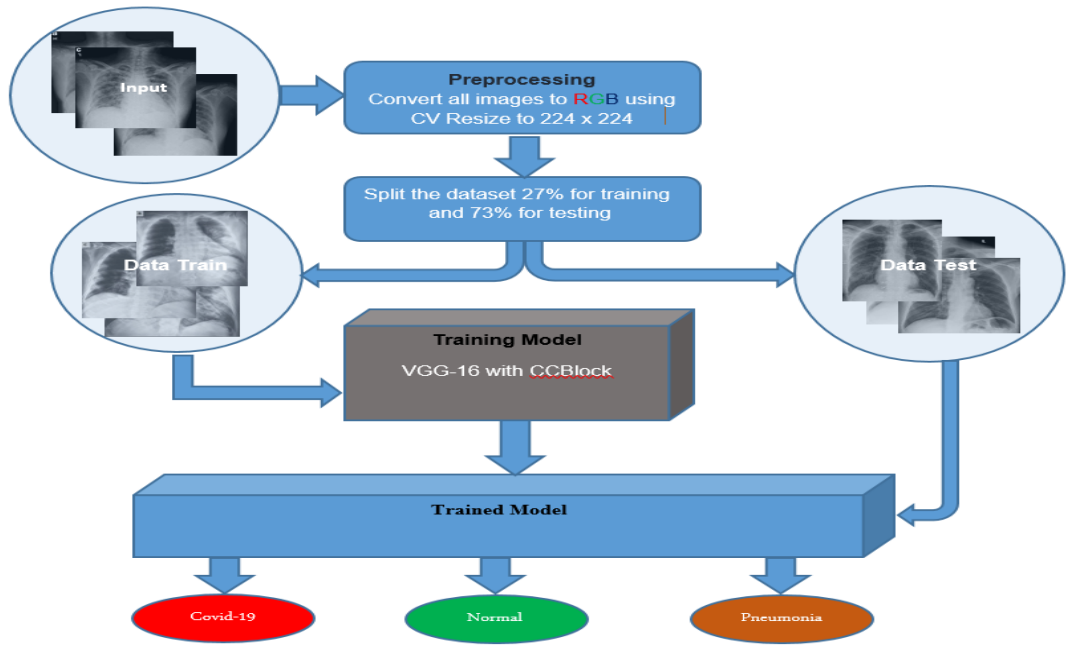

**Fig. 2:** The procedure for the proposed model

## Experiments and Results

In this study, different tests were conducted to diagnose and classify COVID-19 cases through chest radiography. The tests were conducted on two types of databases, the first of which included two categories (COVID-19, Normal), whereas the second included three categories (COVID-19, Normal, and Pneumonia). The dataset was divided into two sections (27% for training data and 73% for test data). To evaluate the efficiency and stability of the proposed model, the tests were repeated five times on both types of data. The optimizer (SGD) was used at a 0.001 learning rate, batch size of 32, the momentum of 0.9, and 30 epochs.

This study was conducted through Python and Keras packages with TensorFlow on an Intel (R) Core (TM) i7-5700 HQ CPU running at 2.70GHz (8 CPUs). Moreover, the experiments were carried out on a computer running on the NVIDIA GTX 970M with 8 GB of GDDR and 16 GB of RAM. The code runtime (disease diagnosis time) was less than one second.

Fig. 3 shows the graph of classification loss as well as the accuracy rates of training and testing phases.

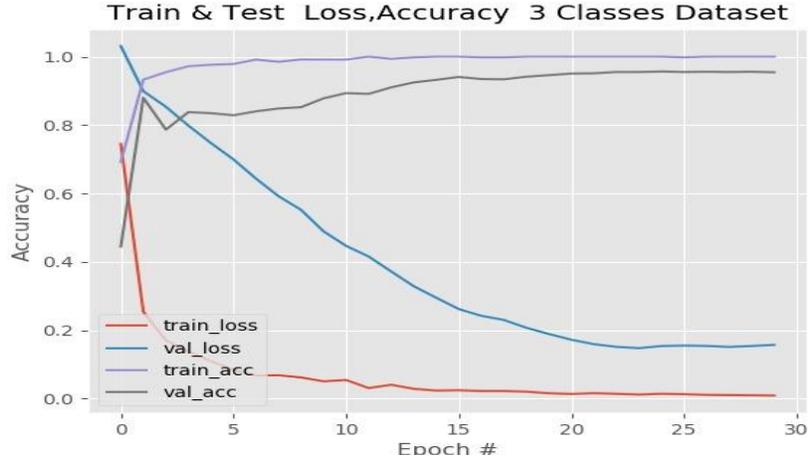

**Fig. 3:** The graph of classification loss and accuracy of the proposed model

According to Fig. 3, training loss decreases rapidly, as the results show an approximate rate of loss of 0.1 during the first five epochs. The rate of loss continued downward until it reached nearly zero after 25 epochs. As for the rate of test losses, its descent was less steep, and this is normal because the data that tested the proposed model were new. Regarding the accuracy scheme, it is clear that the proposed model is generalizable, as there is a slight difference in training accuracy and testing accuracy. This shows the good efficiency of the proposed model *CCBlock*.

To evaluate the proposed model *CCBlock*, a confusion matrix was calculated for each implementation phase (Fig. 4 and Fig. 5). The results showed that the proposed model was characterized by stability and efficiency in diagnosing the COVID-19 for different categories (Normal and Pneumonia). A rate of 98.52% accuracy was reported for the two categories, whereas a rate of 95.34% accuracy was documented for the three categories. Table 3 presents the values of sensitivity, specificity, and accuracy for three categories and five implementation times. Table 4 shows the same values for two categories.

**Table 3:** sensitivity, specificity, and accuracy for three categories

|         | Sensitivity | Specificity | Accuracy |
|---------|-------------|-------------|----------|
| Run1    | 98.21       | 98.94       | 95.21    |
| Run2    | 99.10       | 98.72       | 95.43    |
| Run3    | 99.10       | 99.15       | 95.43    |
| Run4    | 96.85       | 99.36       | 95.13    |
| Run5    | 99.10       | 98.72       | 95.51    |
| **average** | **98.47** | **98.98** | **95.34** |

**Table 4:** sensitivity, specificity, and accuracy for two categories

|         | Sensitivity | Specificity | Accuracy |
|---------|-------------|-------------|----------|
| Run1    | 98.67       | 98.54       | 98.58    |
| Run2    | 98.23       | 98.54       | 98.44    |
| Run3    | 98.67       | 97.70       | 98.01    |
| Run4    | 98.66       | 98.95       | 98.86    |
| Run5    | 98.67       | 98.74       | 98.72    |
| **average** | **98.58** | **98.49** | **98.52** |

According to Table 3, the proposed model was proven to be efficient in diagnosing and differentiating the COVID-19 cases from the other classes (Normal, Pneumonia). The highest rate of accuracy was reported at 95.51%.

However, the average of five implementation times was obtained, and a rate of 95.34% was recorded. To evaluate the efficacy of the proposed model on the diagnosis and classification of the COVID-19, *CCBlock* was tested on the second dataset including the x-rays of patients with the COVID-19 as well as the images of the uninfected people. The highest accuracy was recorded 98.86% for the proposed model; however, the average of five executions was calculated and considered the accuracy of the proposed model. The average accuracy was reported at 95.34%.

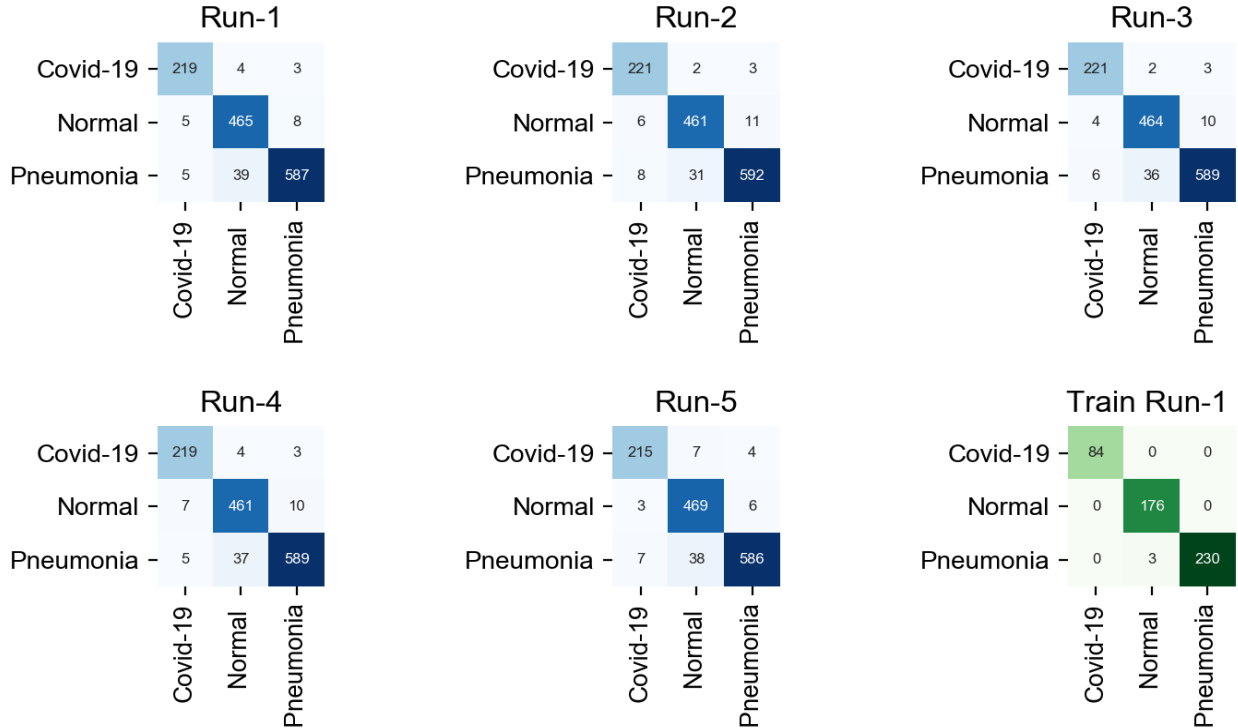

**Fig. 4:** The confusion matrix of the proposed three-class model

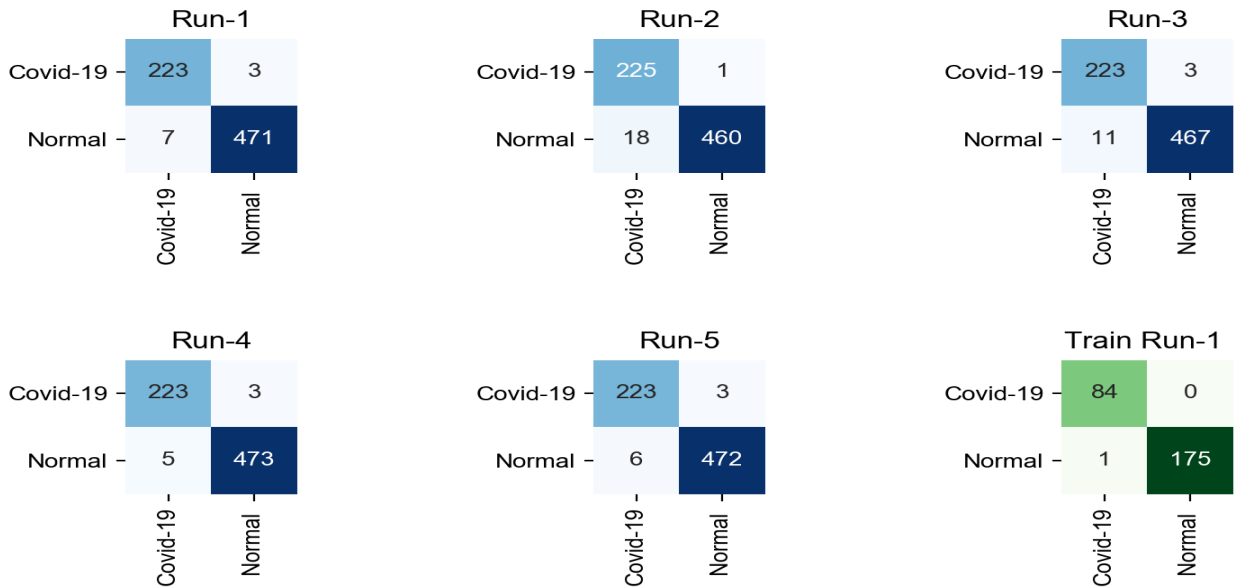

**Fig. 5:** The confusion matrix of the proposed two-class model

Regarding machine learning and classification, in particular, a confusion matrix is an instrument that allows more precise visualization of the algorithm performance, as it shows errors of a classification algorithm for each category in comparison with other categories. The primary diameter of the array represents the classes that were correctly categorized, whereas the other elements represent the data that were incorrectly classified as other categories. Fig.4 shows the confusion matrix for each implementation on the first dataset including three categories (COVID-19, Normal, and Pneumonia). Accordingly, the proposed model is highly capable of diagnosing and distinguishing the COVID-19 from other categories. The highest rate of accuracy was reported by 98% for the COVID-19. The Train Run-1 matrix shows the COVID-19 classification for the first implementation by applying the proposed model on the training data for three categories.

Fig. 5 indicates the confusion matrix of a series of tests conducted on the second dataset including two categories (COVID-19, Normal). This study focused on the COVID-19; therefore, it is preferable to evaluate the efficiency of the proposed model *CCBlock* in diagnosing and distinguishing the COVID-19 cases from those who are not infected. According to the confusion matrix, the proposed model managed to record a diagnostic accuracy of 99.55% for the COVID-19. The Train Run-1 matrix showed the COVID-19 classification for the first implementation by using the proposed model on the training data for two categories.

The previous studies showed that deep neural networks were efficient in diagnosing and distinguishing the COVID-19 cases properly from other categories. However, the proposed model *CCBlock* proved outperformed the previous methods in both cases of two categories and three categories. In fact, it yielded higher accuracy than the previous techniques, as shown in Table 5.

## Discussion

A critical problem that researchers face in the use of deep learning for diagnosis and treatment through medical images is the lack of available data on such tasks. Therefore, researchers tend to use deep learning along with the transfer learning strategy to solve this problem. It is necessary to access sufficient numbers of images or data to train a deep model. This study proposed the VGG-16 + *CCBlock* model in two parts, the first of which (VGG-16) used a transfer learning strategy, whereas the second (*CCBlock*) was trained from the beginning by using the available training data introduced in the *Database*. Due to the harsh conditions of the COVID-19 pandemic, it is impossible to collect sufficient data. Thus, most researchers relied on the COVID-19 dataset available on the Kaggle platform.

The efficiency of the proposed model was evaluated by using two datasets, the first of which included a category of the infected and uninfected people. The accuracy of this dataset was reported at 98.86%. This shows the ability of the proposed model to diagnose the people with the Covid-19 accurately.

To increase the complexity of the issue and test the efficiency of the proposed model incorrect diagnosis, a third category (Pneumonia) was added to the dataset. In fact, pneumonia is a medical condition that affects the respiratory system and lungs in particular. Therefore, the model faces difficulty in distinguishing between these two similar categories (COVID-19, Pneumonia).

As mentioned earlier, the proposed model showed high efficiency in distinguishing between these two categories with high accuracy (95.51%).

The efficiency of the proposed model was also tested in distinguishing between the two categories (COVID-19, Pneumonia) by conducting several tests on the dataset. The accuracy was reported by 98.95%.

Fig. 4 and Fig. 5 show the confusion matrices for the two datasets of two and three classes, in which (FP, FN, TP, TN) were calculated for the two groups. Table 5 displays the results.

The error rates were close and very small for the three classes, a finding which indicates the efficiency of the proposed model in differentiating the three classes perfectly.

**Table 5:** FP, FN, TP, and TN for one of the tests on the two datasets

| Categories | TP | TN | FP | FN |
|---|---|---|---|---|
| **Two categories** | 223 | 473 | 5 | 3 |
| **Three categories** | 221 | 1099 | 10 | 5 |

The proposed model proved its efficiency in diagnosing the research categories, as an accuracy of 98.86% was recorded for the two-category dataset. This finding exceeds the most similar study (mentioned in Table 6) which recorded accuracy of 98.08% on a two-category dataset. However, the number of images used in our study exceeded the number of pictures in [19] as shown in Table 6.

Regarding the three-category dataset, the proposed model also proved its efficiency in distinguishing these categories, especially the COVID-19 and pneumonia. The distinction between these two categories is important because they affect the respiratory system and cause similar phenotypic changes. An accuracy of 95.51% was recorded for this dataset. It exceeds the findings of similar studies mentioned in Table 6.

**Table 6**: Comparison between the proposed model and the previous studies

| Study | Type of Images | Number of Cases | Method Used | Accuracy 2-classes (%) | Accuracy 3-classes (%) |
|---|---|---|---|---|---|
| Ioannis et al. [18] | Chest X-ray | 224 COVID-19(+) 700 Pneumonia 504 Healthy | VGG-19 | - | 93.48 |
| Tulin et al. [19] | Chest X-ray | 125 COVID-19(+) 500 No-Findings  125 COVID-19(+) 500 Pneumonia 500 No-Findings | DarkCovidNet | 98.08 | 87.02 |
| Wang and Wong [20] | Chest X-ray | 53 COVID-19(+) 5526 COVID-19(-) 8066 Healthy | COVID-Net | - | 92.4 |
| Sethy and Behra [23] | Chest X-ray | 127 COVID-19(+) 127 Normal 127 Pneumonia | ResNet50+ SVM | - | 95.33 |
| Zheng et al. [26] | Chest CT | 313 COVID-19(+) 229 COVID-19(-) | UNet+3D Deep Network | 90.8 | - |
| Wang et al. [27] | Chest CT | 195 COVID-19(+) 258 COVID-19(-) | M-Inception | 82.9 | - |
| Xu et al. [28] | Chest CT | 219 COVID-19(+) 224 Viral pneumonia 175 Healthy | ResNet+Location Attention | - | 86.7 |
| Ying et al [29] | Chest CT | 777 COVID-19(+) 708 Healthy | DRE-Net | 86 | - |
| **Proposed Study CCBlock** | Chest X-ray | 310 COVID-19(+) 654 Healthy 864 Pneumonia(virus &bacteria) | VGG-16+CCBlock | 98.86 | 95.51 |

## Conclusion

Deep neural networks have proven to be efficient in diagnosing respiratory diseases through the chest radiography. This study proposed a model for diagnosing the COVID-19 through a transfer learning strategy. The *CCBlock* and VGG-16 were trained by ImageNet. Excellent results were obtained, showing the definitive diagnosis of the COVID-19. The experiments indicated that that the proposed model was highly efficient in diagnosing the COVID-19 with an accuracy of 98.86% for two categories (COVID-19, Normal) and an accuracy of 95.51% for three categories (COVID-19, Normal, Pneumonia).

The results point out that the proposed model can help radiologists make better diagnosis decisions. Despite the high accuracy of computer-aided diagnosis, laboratory tests such as PCR cannot be dispensed with. However, these results can be employed to assist and support laboratory results. In the future, if sufficient data are available, deep neural networks can be trained from the beginning and can yield better results without any need for a transfer learning strategy.

## Acknowledgment

It was impossible to complete this work of research without the participation and assistance of so many people whose names may not all be mentioned. Their contributions are sincerely appreciated and gratefully acknowledged. However, the group would like to express their deepest appreciation and indebtedness particularly to Dr. Reza Monsefi, Dr. Abedin Vahedian, Dr. Kamaledin Ghiasi-Shirazi, and Dr. Ahad Harati for their endless support, kindness, and understanding spirits during our case presentation. Our utmost gratitude goes to all relatives, friends, and all of those who shared their support either morally and physically. Above all, to God Almighty, the author of knowledge and wisdom, for his countless love. Thank you.

**Conflict of Interest:** The authors declare that they have no conflict of interest.

## References

1. Na Zhu, Dingyu Zhang, Wenling Wang, et al. A Novel Coronavirus from Patients with Pneumonia in China, 2019; New England Journal of Medicine. 2020; doi: 10.1056/NEJMoa2001017.
2. Qun Li, Xuhua Guan, Peng Wu, et al. Early Transmission Dynamics in Wuhan, China, of Novel Coronavirus–Infected Pneumonia. New England Journal of Medicine. 2020; doi: 10.1056/NEJMoa2001316.
3. Jon Cohen ,Dennis Normile. New SARS-like virus in China triggers alarm. Science. 2020; doi: 10.1126/science.367.6475.234
4. Victor M Corman, Olfert Landt, Marco Kaiser, et al. Detection of 2019 novel coronavirus (2019-nCoV) by real-time RT-PCR. Eurosurveillance. 2020; doi:10.2807/1560-7917
5. Paules C, Marston H, Fauci A. Coronavirus Infections—More Than Just the Common Cold. JAMA. 2020; doi:10.1001/jama.2020.0757.
6. Sohrabi C, Alsafi Z, O'Neill N, et al. World Health Organization declares global emergency: A review of the 2019 novel coronavirus (COVID-19). International Journal of Surgery. 2020; doi:10.1016/j.ijsu.2020.02.034.
7. Coronavirus [https://www.who.int/health-topics/coronavirus]. Accessed 25 May 2020
8. Huang C, Wang Y, Li X, et al. Clinical features of patients infected with 2019 novel coronavirus in Wuhan, China. The Lancet. 2020; doi:10.1016/S0140-6736(20)30183-5.
9. Mahase E. Coronavirus: covid-19 has killed more people than SARS and MERS combined, despite lower case fatality rate. BMJ. 2020; doi:10.1136/bmj.m641.
10. Wang W, Xu Y, Gao R, et al. Detection of SARS-CoV-2 in Different Types of Clinical Specimens. JAMA. 2020; doi:10.1001/jama.2020.3786.
11. Ng M, Lee E, Yang J, et al. Imaging Profile of the COVID-19 Infection: Radiologic Findings and Literature Review. Radiology: Cardiothoracic Imaging. 2020; doi:10.1148/ryct.2020200034.

12. Ai T, Yang Z, Hou H, et al. Correlation of Chest CT and RT-PCR Testing in Coronavirus Disease 2019 (COVID-19) in China: A Report of 1014 Cases. Radiology. 2020; doi:10.1148/radiol.2020200642.
13. Kholiavchenko M, Sirazitdinov I, Kubrak K, et al. Contour-aware multi-label chest X-ray organ segmentation. International Journal of Computer Assisted Radiology and Surgery. 2020; doi: 10.1007/s11548-019-02115-9.
14. Sakshi Patel, Bharath K P, Rajesh Kumar Muthu. Medical Image Enhancement Using Histogram Processing and Feature Extraction for Cancer Classification. 2020; [https://arxiv.org/abs/2003.06615]
15. Celik Y, Talo M, Yildirim O, et al. Automated invasive ductal carcinoma detection based using deep transfer learning with whole-slide images. Pattern Recognition Letters 2020; doi:10.1016/j.patrec.2020.03.011.
16. Talo M, Yildirim O, Baloglu U, et al. Convolutional neural networks for multi-class brain disease detection using MRI images. Computerized Medical Imaging and Graphics. 2019; doi:10.1016/j.compmedimag.2019.101673.
17. Pranav Rajpurkar, Jeremy Irvin, Kaylie Zhu, et al. CheXNet: Radiologist-Level Pneumonia Detection on Chest X-Rays with Deep Learning. 2017; [https://arxiv.org/abs/1711.05225]
18. Apostolopoulos I, Mpesiana T: Covid-19: automatic detection from X-ray images utilizing transfer learning with convolutional neural networks. Physical and Engineering Sciences in Medicine. 2020; doi:10.1007/s13246-020-00865-4.
19. Ozturk T, Talo M, Yildirim E, et al. Automated detection of COVID-19 cases using deep neural networks with X-ray images. Computers in Biology and Medicine. 2020; doi:10.1016/j.compbiomed.2020.103792.
20. Linda Wang, Alexander Wong. COVID-Net: A Tailored Deep Convolutional Neural Network Design for Detection of COVID-19 Cases from Chest X-Ray Images. 2020; [https://arxiv.org/abs/2003.09871]
21. Ezz El-Din Hemdan, Marwa A. Shouman, Mohamed Esmail Karar. COVIDX-Net: A Framework of Deep Learning Classifiers to Diagnose COVID-19 in X-Ray Images. 2020; [https://arxiv.org/abs/2003.11055]
22. Ali Narin, Ceren Kaya, Ziynet Pamuk. Automatic Detection of Coronavirus Disease (COVID-19) Using X-ray Images and Deep Convolutional Neural Networks. 2020; [https://arxiv.org/abs/2003.10849]
23. Prabira Kumar Sethy, Santi Kumari Behera. Detection of coronavirus disease (covid-19) based on deep features and support vector machine. 2020 [https://doi.org/10.33889/IJMEMS.2020.5.4.052].
24. Pedro R. A. S. Bassi, Romis Attux. A Deep Convolutional Neural Network for COVID-19 Detection Using Chest X-Rays. 2020; [https://arxiv.org/abs/2005.01578].
25. Li L, Qin L, Xu Z, et al. Artificial Intelligence Distinguishes COVID-19 from Community Acquired Pneumonia on Chest CT. Radiology. 2020; doi:10.1148/radiol.2020200905.
26. Chuansheng Zheng, Xianbo Deng, Qing Fu, et al. Deep Learning-based Detection for COVID-19 from Chest CT using Weak Label. 2020; [https://www.medrxiv.org/content/10.1101/2020.03.12.20027185v2]
27. Shuai Wang, Bo Kang, Jinlu Ma, et al. A deep learning algorithm using CT images to screen for Corona Virus Disease (COVID-19). 2020; [https://www.medrxiv.org/content/10.1101/2020.02.14.20023028v5]
28. Xiaowei Xu, Xiangao Jiang, Chunlian Ma, et al. Deep Learning System to Screen Coronavirus Disease 2019 Pneumonia. 2020; [https://arxiv.org/abs/2002.09334]
29. Ying Song, Shuangjia Zheng, Liang Li, et al. Deep learning Enables Accurate Diagnosis of Novel Coronavirus (COVID-19) with CT images. 2020; [https://www.medrxiv.org/content/10.1101/2020.02.23.20026930v1]
30. Joseph Paul Cohen, Paul Morrison, Lan Dao. COVID-19 image data. 2020; [https://github.com/ieee8023/covid-chestxray-dataset]. Accessed 5 May 2020
31. Larxel. COVID-19 X rays. 2020; [https://www.kaggle.com/andrewmvd/convid19-x-rays]. Accessed 5 May 2020
32. Paul Mooney. Chest X-Ray Images (Pneumonia). 2017; [https://www.kaggle.com/paultimothymooney/chest-xray-pneumonia]. Accessed 5 May 2020
33. Weiss K, Khoshgoftaar T, Wang D: A survey of transfer learning. Journal of Big Data. 2016; doi:10.1186/s40537-016-0043-6.
34. Minyoung Huh, Pulkit Agrawal, Alexei A. Efros. What makes ImageNet good for transfer learning?. 2016; [https://arxiv.org/abs/1608.08614]
35. Russakovsky O, Deng J, Su H, et al. ImageNet Large Scale Visual Recognition Challenge. International Journal of Computer Vision. 2015; doi:10.1007/s11263-015-0816-y.
36. Simonyan K, Zisserman A. Very Deep Convolutional Networks for Large-Scale Image Recognition. In ICLR 2015; 2015.